\documentclass{sig-alternate-05-2015}
\usepackage{subcaption}

\def\sharedaffiliation{%
\end{tabular}
\centering
\begin{tabular}{p{\linewidth}}\centering}
\begin{document}

\setcopyright{acmcopyright}

\doi{10.475/123_4}

\isbn{123-4567-24-567/08/06}

\conferenceinfo{INTIS'2019}{December 20--21, 2019, Tangia, Morocco}

\acmPrice{\$15.00}

%
\conferenceinfo{INTIS}{'2019 Tangier, Morocco}

\title{Including Images into Message Veracity \\ Assessment in Social Media}
%
%
%
%
%

\numberofauthors{4} 
%
\author{
%
%
\alignauthor Abderrazek Azri\\
\email{a.azri@univ-lyon2.fr}
\alignauthor C\'ecile Favre\\
\email{cecile.favre@univ-lyon2.fr}
\and\alignauthor Nouria Harbi\\
\email{nouria.harbi@univ-lyon2.fr}
\alignauthor J\'er\^ome Darmont\\
\email{jerome.darmont@univ-lyon2.fr}\\
\sharedaffiliation
\affaddr{University of Lyon, Lyon 2, ERIC EA 3083}  \\
\affaddr{5 avenue Pierre Mend\`es France,}   \\
\affaddr{F69676 Bron Cedex, France}  
}

\maketitle
\begin{abstract} 
The extensive use of social media in the diffusion of information has also laid a fertile ground for the spread of rumors, which could significantly affect the credibility of social media. An ever-increasing number of users post news including, in addition to text, multimedia data such as images and videos. Yet, such multimedia content is easily editable due to the broad availability of simple and effective image and video processing tools. The problem of assessing the veracity of social network posts has attracted a lot of attention from researchers in recent years. However, almost all previous works have focused on analyzing textual contents to determine veracity, while visual contents, and more particularly images, remains ignored or little exploited in the literature. In this position paper, we propose a framework that explores two novel ways to assess the veracity of messages published on social networks by analyzing the credibility of both their textual and visual contents. 
\end{abstract}

%
%


%
%

%
%



\section{Introduction} 

With the rise of social media platforms, information is generated and propagated at an unprecedented rate. The simplicity of the sharing process has led to a large volume of news content spreading over social networks and reaching vast numbers of users in a short time. 

The convenience and openness of microblogs have also fostered the spread of rumors, which have become a serious public concern recently. The spread of rumors is also facilitated by the availability of low-cost and straightforward multimedia content processing tools, such that everyone can easily edit the content of an image or video \cite{lago2019visual}. Therefore, the automatic assessment of information credibility on social networks becomes a mandatory requirement to limit the propagation of rumors.

On a microblogging platform such as Twitter, users publish short messages (i.e., tweets) that may contain, in addition to ``classical" text, tags such as URLs, hashtags or references to other users. Moreover, the message can be accompanied by multimedia content such as an image or video. Consequently, the textual and visual content that forms the message should be considered together. Moreover, the message's veracity depends on the veracity of each of these components.


	
	
%
		

	
	

Most existing approaches for automatic rumor detection are based only on the textual content of messages to predict the veracity of online content. Unlike these approaches, we propose in this position paper a framework that explores the veracity of messages from social networks by analyzing both their textual and visual content in a single process.

For this purpose, we propose two techniques to achieve this challenging task. The first is based on the extraction of appropriate features from the text and useful statistical and visual features from the image. These features are merged to train a supervised classifier to assess the veracity of messages. 

The second technique involves a verification system composing two classification tiers. The first classification tier determines the veracity of the textual content based on textual feature extraction; the second classification tier determines the veracity of the image by adopting a forgery detection method. Finally, scores returned from both classifiers are fused to provide the final decision about message veracity.

The rest of this paper is structured as follows. We present the state of the art in Section~\ref{sec:relatedworks}. We detail the definition and categories of fake images circulating on social media in Section~\ref{sec:fakeimage}. We describe the framework we  propose in Section~\ref{sec:framework}. We discuss our framework in Section~\ref{sec:Discussion}. We conclude this paper and present perspectives for our future work in Section~\ref{sec:conclusionperspective}. 

\section{related works} 
\label{sec:relatedworks}

We review related work in two main areas: rumor analysis and images forgery detection. Because these are two extensive research areas, we just provide an outline of the research that is most closely related to ours.

\subsection{Rumor Analysis in Social Media}

Almost existing studies solve the rumor detection problem in feature-based approaches and supervised machine learning scheme. Features from text content \cite{castillo2011information,kwon2013prominent}, users \cite{morris2012tweeting}, and propagation patterns \cite{wu2015false}, are extracted to train a classifier on labeled training data. Some recent works further
improve the classification result with graph-based optimization methods \cite{gupta2012evaluating,jin2014news}.

To represent high-level abstract semantics in the rumor detection task, the approaches of \cite{jin2016news,wu2015false} use features based on topic modeling (LDA). To overcome the limitation of these manually crafted features, \textit{Ma et al.} \cite{ma2016detecting} represents tweets in an event with deep neural networks by using recurrent neural networks (RNN), to learn the representation of tweets in a time series.  

Only a few recent studies aim to verify the credibility of multimedia content in addition to text. In \cite{gupta2012evaluating,wu2015false}, some basic features are provided for images attached in tweets. Text and image features are extracted to automatically predict the veracity of a tweet that shares multimedia content in \cite{boididou2015verifying}. 

Finally, \textit{Jin et al.} \cite{jin2017multimodal} proposes an RNN with the attention mechanism to fuse features from the text, image, and social context for detecting rumors on microblogs. However, the proposed system in \cite{jin2017multimodal} has several limitations. First, this system does not explain the final classification, this problem is one of the weak points in the use of neural networks. Also, the system as it stands, cannot handle the detection of information misuse, this corresponds to a bad association of an image and a text representing two different contexts.

\subsection{Image Forgery Detection}

Creating forged images by manipulating the original image content is called digital image forgery. This field that intends to verify the authenticity of images has been developed significantly against the problem of image forgeries in many domains like legal services, medical images, and forensics. 

There are two main ways to detect alteration: classification and localization. In the classification detection method, the output is binary, whether the image is authentic or forged. On the other hand, the localization method not only detects whether an image is authentic or falsified, but also gives the regions that are manipulated in the image if it's falsified \cite{qureshi2015bibliography, warif2016copy}.

Image forgery detection methods can be divided into two types: active and passive. An active forgery detection technique uses a known authentication code embedded into the image content such as digital watermarking \cite{lin1998generating} or digital signatures \cite{shieh2006semi}. Passive forgery detection technique is the process of authenticating images with no requirement of prior information, just the image itself \cite{asghar2017copy, birajdar2013digital}. 

In passive forgery detection, if a forged image involves parts of more than one image, then resultant image is called a spliced image; if the forged image involves parts of the same image, it is called a copy-move forged image.

\section{Fake Images in Social Media}
\label{sec:fakeimage}

The content of a message on social media is usually linked to an event or a context. For this purpose, the definition of a fake image leads us first of all to define the event.

In literature, the definition of an event can be heterogeneous even though it shares a common characteristic; events are in general said to occur, or happen, meaning that they are entities that unfold over time and/or space \cite{scherp2009f, wang2007eventory}. 

A fake image is defined as any image attached to a message that does not accurately represent the event to which it refers \cite{boididou2017learning}.

Following the definition of the event, a fake image is defined as either miscontextualized or tampered. A miscontextualized image presents inconsistencies with at least one aspect of the event, for instance, temporal or geographical misplacement. In other words, it is a bad association of an image and a text representing two different contexts.

\begin{figure*}[ht]
	\centering
	\begin{minipage}[l]{1.0\columnwidth}%
			\centering
			\includegraphics[width=0.9\columnwidth]{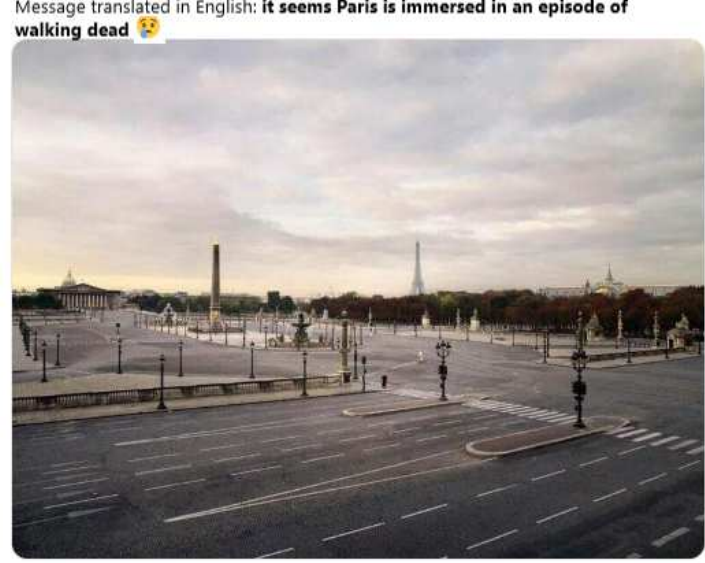}
	         \subcaption{}	
	\end{minipage}%
	\hfill
	\begin{minipage}[r]{1.0\columnwidth}%
		\centering
		\includegraphics[width=0.9\columnwidth]{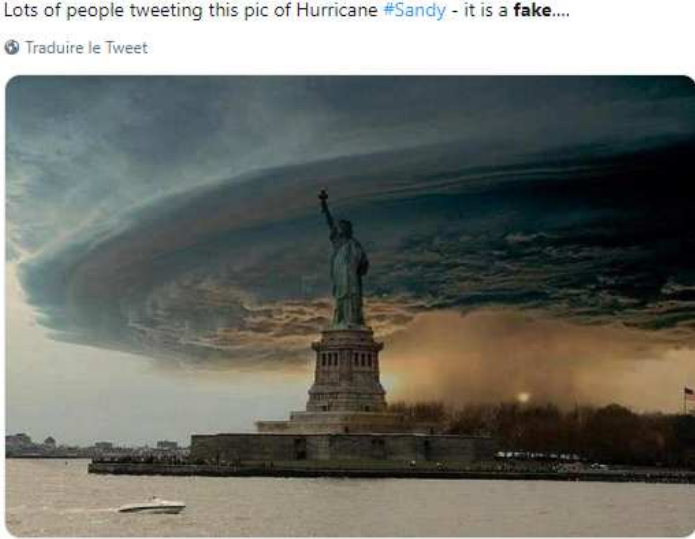}
		\subcaption{}
	\end{minipage}

	\caption{Different types of fake images: (a) miscontextulized image; (b)  Digitally manipulated image.}\label{Fig 2}
\end{figure*}

Figure~\ref{Fig 2}(a) shows a miscontextulized image that seems to
show Paris the day after the attacks of November 2015.
However, by searching on Google Image, we can find the origin
of the image that dates back to 2008. While the image used to show the hurricane sandy in Figure~\ref{Fig 2}(b) is deliberately tampered .

\section{The Proposed Framework} 
\label{sec:framework}

The key idea of our proposal is to exploit both textual and visual content of messages to assess the veracity of entire online content by studying the possibilities of text and image processing in a single framework.

\begin{figure*}[h]
	\centering
	\includegraphics[width=.65\textwidth]{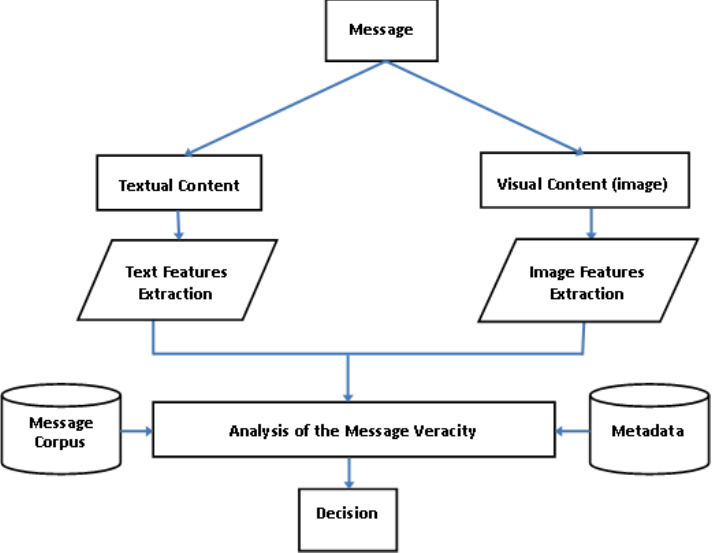}
	\caption{Overview of the proposed framework.}\label{Fig3} 
\end{figure*}

Figure~\ref{Fig3} presents an overview of the main components of the proposed framework.	

\subsection{First Technique}

The detection of the veracity of messages is formalized in this technique as a classification problem, and it mainly contains two phases: features extraction and model training.

In this method, several useful features are extracted from the message text and the attached image. After features extraction, "traditional" classifiers are built, based on these features to identify messages veracity as real or fake.

\subsubsection{\textbf{Features Extraction}}
\paragraph{\textbf{Textual Features Extraction}}

Features come from two main aspects of false information or rumor: content and social context in social media, this because the content is associated with a certain social context during spreading of rumor.

Content features are the features extracted from the text. The social context reflects the relationship among different users and describes the propagating process of a rumor, so social context features are the features extracted from the user behavior and the propagation network. These features are suitable to capture the characteristics of rumors.

Depending on their nature, we distinguish three features categories: message content features, user content features, and propagation properties features. 

\begin{enumerate}
	
\item \textbf{Message Content Features.} They take into account information about the content of the messages themselves; this information may or may not be related to social media. The information that is not related to social media includes the presence of exclamation or question marks, the presence of positive or negative sentiment words, and the size of the message. The information that is linked to social media includes the presence of hashtags, URLs, the fact that the message is original or a re-tweet.
	
\item \textbf{User Content Features.} They are the attributes of the user who posted the message, e.g., we assume properties such as the number of friends, followers, the number of tweets the user has authored in the past or registration age. 

\item \textbf{Propagation Properties Features.} They take into account information about the propagation tree that can be built from tweet and re-tweets of a message, such as the depth of the re-tweet tree.

\end{enumerate}

We propose a set of features to characterize messages in our dataset. Many of these features have been studied in previous works \cite{agichtein2008finding,boididou2017learning,qazvinian2011rumor,ratkiewicz2010detecting}. 

Patterns of these categories features for rumor messages are obviously different from that of "normal" ones. Table~\ref{table 1} presents a list of features produced for each message.  

\begin{table*}[ht]
	\centering
	\caption{List of Features Extracted from Each Message.}\label{table 1}
	\begin{tabular}{|l|l|}
		\hline
		Feature category &  Description \\
		\hline
		& Length of the tweet \\
		& Number of Words \\
		& Contains Question Mark ? \\
		& Contains Exclamation Mark !\\
		& Contains Happy Emoticon \\
		& Contains Sad Emoticon \\
		& Contains a personal pronoun in 1st, 2nd, or 3rd person \\
		& Number of uppercase characters\\
		Message  & Number of negative sentiment words \\
		& Number of positive sentiment words \\
		& Number of mentions \\
		& The date in which this tweet was written\\
		& Number of hashtags \\
		& Number of URLs \\
		& Retweet count \\  \hline
		& Number of Friends\\
		& Number of Followers \\
		& Follower-Friend Ratio \\
		User& Age of user account \\
		& Is a verified user\\
		& User has a URL\\
		& The number of tweets at posting time \\ \hline
		& The depth of a propagation tree\\
		Propagation & The max. size of a level in the propagation tree \\
		& The degree of the root in a propagation tree\\ 
		\hline
		
	\end{tabular}
\end{table*}

\paragraph{\textbf{Image Features Extraction}}

To make the difference between real and fake images, the authors of \cite{jin2016image} find that, to describe false information, fake images tend to be eye-catching, and visually striking in contrast to real ones. Also, images in the real news are much denser than those in the fake news. Hence, images in fake and real news have visually and statistically distinctive patterns. Therefore, images have an important impact on detecting fake news in microblogs.

For this purpose, we propose a set of features to characterize images, which are divided into two categories. 

\begin{enumerate}
	
 \item \textbf{Statistical Features.} Similar to the statistical features of textual content, some basic statistics of images proved to be distinctive in separating rumors and non-rumors as to mark the occurrence of images in rumor messages, or popular images that gain more retweeting and comments than others.

\item \textbf{Visual Features.} Extracted from Image quality assessment (IQA) field. IQA aims to quantitatively represent the human perception of quality. These metrics are commonly used to analyze the performance of algorithms in different fields of computer vision like image compression, image transmission, and image processing \cite{maitre2017photon}.

IQA is mainly divided into two areas of research:~(1) Full-Reference evaluation and~(2) No-reference evaluation. Full-Reference algorithms compare the input image against a pristine reference image with no distortion. In no-reference algorithms, the only input is the image whose quality we want to measure, these algorithms compare statistical features of the input image against a set of features derived from an image database.

Since in our case, we don't have the original version of the posted image; therefore, the approach that is fitting for our context is no-reference IQA metrics. 

For this purpose, we use three No-reference algorithms demonstrated to be highly efficient:  Blind/Referenceless Image Spatial QUality Evaluator (BRISQUE) \cite{mittal2011blind}, Naturalness Image Quality Evaluator (NIQE) \cite{mittal2012making}, and Perception based Image Quality Evaluator (PIQE)\cite{venkatanath2015blind}.

\end{enumerate}

Table~\ref{table 2} presents a list of some features produced for each image.  

\begin{table*}[ht]
	\centering
	\caption{List of Features Extracted from Each Image.}\label{table 2}
	\begin{tabular}{|l|l|}
		\hline
		Feature category &  Description \\
		\hline
		& BRISQUE score \\
		Visual  & NIQE Score \\
		& PIQE Score \\  \hline
		  & The number of all images in a news event \\
		Statistical& The ratio of the image-tweets in all tweets\\
		& The ratio of image number to tweet number\\
		
		\hline
		
	\end{tabular}
\end{table*}

\subsubsection{\textbf{Model Training}}

Extracted features are normalized and concatenated, and a classifier model will be trained using the features. A large amount of supervised model can be used such as Support Vector Machine, Random Forest, Decision Tree, and Na\"ive Bayes. At the end of this step, messages are labeled true or false.

\subsection{Second Technique}

In this technique, we propose a verification system based on three main steps.

First, we explore the veracity of textual content by adopting the approach followed in the first proposition, using only the features listed in Table~\ref{table 1} and a classifier whose output is the probability of the veracity of textual content of the message, e.g., logistic regression.

Secondly, we explore the veracity of visual content by adopting an image forgery detection method. Almost of the existing forgery detection approaches extract representative and relevant features from an image first, then a suitable classifier is trained and modeled by using the features, and finally, classification performed by using the trained model. 

In the field of image forgery detection, no method can detect all forms of falsification, because each technique is designed to identify specific traces of manipulation based on its own assumption; it is, therefore, judicious to fuse multiple outputs from many forgery techniques \cite{fontani2013framework}. Basing on this acknowledgment, we propose to combine numerous classical image forgery techniques on the expectation to detect various image manipulations. 

The fusion of these techniques makes it possible to detect more precisely the region that is more likely to contain falsification. Then, relevant statistics are extracted from this region to generate a features vector which will be used to train a classifier, for example, random forests, that provide the probability of image veracity.

Finally, scores returned from the two classifiers are fused to give the final decision.

The final probability of veracity of the whole message is defined as a linear combination of the probability outputs from these two classifiers, as follows:    

\begin{equation}
\label{eq:equation1}
Pr(Message)= \omega_1\times Pr(Txt_{content}) + \omega_2\times Pr(Img_{content})
\end{equation}

Where, $Pr(Txt_{content})$ and $Pr(Img_{content})$ are respectively the probabilities to be fake for the textual content and the image. The parameters $\omega_1$ and $\omega_2$ are two percentages, where there sum is 100\%. We make all possible combinations of $\omega_1$ and $\omega_2$ to evaluate which values are most appropriate to evaluate the authenticity of messages.

\begin{figure*}[ht]
	\centering
	\includegraphics[width=.65\textwidth]{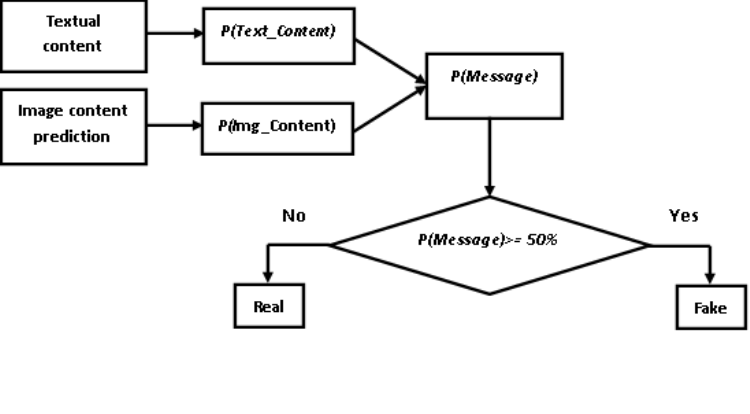}
	\caption{Overview of the second technique.}\label{Fig4} 
\end{figure*}

The linear combination in equation~\eqref{eq:equation1} represents the probability of message to be fake: $Pr(Message)$. If this probability is higher than 50\%, then the image is classified as fake, otherwise as real, as can be seen in Figure~\ref{Fig4}. 

\section{Discussion}
\label{sec:Discussion}

From the analysis of our framework, we can make the following observations and proposals.

The task of distinguishing the credibility of images is challenging, as images can be misleading in many ways. Indeed, in case of miscontextualized images, the vast majority of information misuse requires information from outside the message because it is impossible to detect misuse of information by taking into account only the text and image of a publication.

To overcome this difficulty, and based on the idea that image misused from its context has necessarily got an original context, we propose to search this context on a search engine like Google image\footnote{https ://images.google.com/} or TinEye\footnote{https ://www.tineye.com/} using this same image. To compare the context of our message with those returned by search engines, we can compare the text of the message with the content of found pages.

In the case where the contexts are not identical, there is then misuse of context either in the request message or in the analyzed page. In this situation, the comparison of the publication dates could make the difference between these cases.

Another possibility is to compare the textual content with the image by extracting the most representative words from the text message \cite{drouin2003term}. For the image, we can use a system for searching objects and persons in an image \cite{lavi2018survey} which is an active domain, or a bag of visual words technique. Then by comparing the two contents, we can know if, for example, an image represents a person or an object that is not described in the text.	

In the field of image forgery detection, the assessment of image veracity depends on the ability of detection methods used to  locate alterations. A limitation of these methods is estimating the ability of a modification to make the image fake. Indeed, some modifications will not be intended to mislead the user, such as changing the color of the image. On the other hand, changing a person's face or adding or deleting objects in an image completely changes the user's understanding of the image and therefore its meaning. This aspect related to the threshold for which an image is considered as fake or not has to be further investigated to not send unnecessary warnings to a user who uses an alert system. 

\section{Conclusion and Perspectives} 
\label{sec:conclusionperspective}

Existing approaches for news verification on microblogs ignore image content, which is nonetheless very important,  in messages such as tweets. In this position paper, we present and discuss a framework to analyze the veracity of messages in social media. 

The  framework we propose explores two ways to assess the veracity of messages published on social networks, by analysing both their textual and visual content. The first way is based on the extraction of appropriate features from the text and statistical and visual features from the image. These features are combined to train a supervised classifier to assess the veracity of messages. The second approach presents a verification system with two levels of classification. The first level is used to determine the veracity of textual content; the second determines the veracity of the image by applying a forgery detection method. The returned scores are taken into account to evaluate the final decision on message veracity.

As perspectives for future work, we plan to implement and test our framework with both approaches. However, there are very few standard multimedia rumor detection datasets available. Thus, in addition to the dataset from the MediaEval Verifying Multimedia Use benchmark \cite{boididou2014challenges}, we intend to build a new dataset for the task. Moreover, to communicate the verification process to end-users, we propose to add a visualization module to preview the results.

\bibliographystyle{abbrv}
\bibliography{sigproc}  %
\end{document}